\documentclass[a4paper]{article}

\usepackage{INTERSPEECH2022}
\usepackage{caption,subcaption}
\usepackage[hidelinks]{hyperref}
\usepackage{tikz}

\usetikzlibrary{positioning}
\tikzstyle{block} = [draw, fill=none, rectangle, 
    minimum height=3em, minimum width=6em]
\definecolor{mygreen}{RGB}{2,179,2}
\definecolor{myblue}{RGB}{155,155,255}

\newcommand{\spkrA}{a}
\newcommand{\spkrB}{b}

\def\x{{\mathbf x}}
\def\y{{\mathbf y}}
\def\w{{\mathbf v}}

\def\p{{\bar p}}

\DeclareMathOperator*{\argmax}{arg\,max}

\DeclareMathOperator*{\dnn}{NN}

\title{Revisiting joint decoding based multi-talker speech recognition with DNN acoustic model}
\name{
    Martin Kocour$^{\star}$\qquad
    Kate\v{r}ina \v{Z}mol\'{i}kov\'{a}$^{\star}$\qquad
    Lucas Ondel$^{\star 1}$\qquad\sthanks{Lucas Ondel is now affiliated with LISN, CNRS, Université Paris-Saclay, France.}
    J\'{a}n \v{S}vec$^{\star}$\\
    Marc Delcroix$^{\dagger}$\qquad
    Tsubasa Ochiai$^{\dagger}$\qquad
    Luk\'{a}\v{s} Burget$^{\star}$\qquad
    Jan ``Honza'' \v{C}ernock\'{y}$^{\star}$
    \thanks{The work was partly supported by European Union’s Horizon 2020 projects No. 864702 - ATCO2 and No. 884287 HAAWAII, and by Czech Ministry of Education, Youth and Sports from project no. LTAIN19087. Part of high-performance computation run on IT4I supercomputer and was supported by the Ministry of Education, Youth and Sports of the Czech Republic from the Large Infrastructures for Research, Experimental Development and Innovations project e-Infrastructure CZ – LM2018140.}
}
\address{
    $^{\star}$ Brno University of Technology, Faculty of Information Technology, Speech@FIT, Czechia \\
    $^{\dagger}$ NTT Corporation, Japan
}
\email{ikocour@fit.vut.cz}

\begin{document}

\maketitle
\begin{abstract}
In typical multi-talker speech recognition systems, a neural network-based acoustic model predicts senone state posteriors for each speaker. These are later used by a single-talker decoder which is applied on each speaker-specific output stream separately. In this work, we argue that such a scheme is sub-optimal and propose a principled solution that decodes all speakers jointly. We modify the acoustic model to predict joint state posteriors for all speakers, enabling the network to express uncertainty about the attribution of parts of the speech signal to the speakers. We employ a joint decoder that can make use of this uncertainty together with higher-level language information. For this, we revisit decoding algorithms used in factorial generative models in early multi-talker speech recognition systems. In contrast with these early works, we replace the GMM acoustic model with DNN, which provides greater modeling power and simplifies part of the inference.  We demonstrate the advantage of joint decoding in proof of concept experiments on a mixed-TIDIGITS dataset.
\end{abstract}

%
%

\begin{figure*}[!ht]
    \centering
    \begin{subfigure}[b]{0.48\textwidth}
        \centering
        \begin{tikzpicture}[auto]
            \node [block, inner sep=0em, fill opacity=0.8] (mix) {\includegraphics[width=6em]{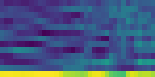}};
            \node[draw=none, below=0 of mix]{\footnotesize $t$};
            \node [draw=none, above=0.2 of mix, fill=none, text width=2.5cm, align=center, inner sep=0cm] (mixlab) {mixed speech $\y$};
            \node[block, right=0.7cm of mix, text width=1.5cm, align=center,rounded corners=0.1cm, inner sep=0.5em] (am) {Acoustic model DNN};
            \node[block, above right=0cm and 0.9cm of am, inner sep=0.1em](out1) {\includegraphics[width=6em]{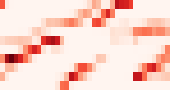}};
            \node[draw=none, below=0 of out1]{\footnotesize $t$};
            \node[draw=none, right=0 of out1]{\footnotesize $v_t^{\spkrA}$};
            \node[draw=none, above=0 of out1, text width=4.9cm, align=center]{speakerA output\\ $p(v_t^\spkrA | \y)$};
            \node[block, below right=0cm and 0.9cm of am, inner sep=0.1em](out2) {\includegraphics[width=6em]{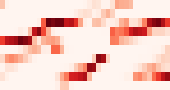}};
            \node[draw=none, below=0 of out2]{\footnotesize $t$};
            \node[draw=none, right=0 of out2]{\footnotesize $v_t^{\spkrB}$};
            \node[draw=none, above=0 of out2, text width=4.9cm, align=center]{speakerB output \\$p(v_t^{\spkrB} | \y)$};
            
            \draw [->] (mix) -- (am);
            \draw [->] (am.east) -- (out1.west);
            \draw [->] (am.east) -- (out2.west);
        \end{tikzpicture}
        \caption{Model with separate output for each speaker.}
        \label{fig:conventional_asr}
     \end{subfigure}
     \hfill
     \begin{subfigure}[b]{0.48\textwidth}
        \centering
        \begin{tikzpicture}[auto]
            \node [block, inner sep=0em, fill opacity=0.8] (mix) {\includegraphics[width=6em]{figs/mfccs.png}};
            \node[draw=none, below=0 of mix]{\footnotesize $t$};
            \node [draw=none, above=0.2 of mix, fill=none, text width=2.5cm, align=center, inner sep=0cm] (mixlab) {mixed speech $\y$};
            \node[block, right=0.7cm of mix, text width=1.5cm, align=center,rounded corners=0.1cm, inner sep=0.5em] (am) {Acoustic model DNN};
            \node[draw=none, right=0.9cm of am, inner sep=0.1em, minimum width=3em, fill=white](out1) {\includegraphics[width=3em]{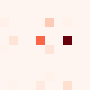}};
            \node[block,above right=-1.6em and -1.6em of out1, inner sep=0.1em, minimum width=3em, fill=white](out6) {\includegraphics[width=3em]{figs/posts_joint.png}};
            \node[block,above right=-2em and -2em of out1, inner sep=0.1em, minimum width=3em, fill=white](out5) {\includegraphics[width=3em]{figs/posts_joint.png}};
            \node[block, above right=-2.4em and -2.4em of out1, inner sep=0.1em, minimum width=3em, fill=white](out2) {\includegraphics[width=3em]{figs/posts_joint.png}};
            \node[block,above right=-2.8em and -2.8em of out1, inner sep=0.1em, minimum width=3em, fill=white](out3) {\includegraphics[width=3em]{figs/posts_joint.png}};
            \node[block, right=0.9cm of am, inner sep=0.1em, minimum width=3em, fill=white](out4) {\includegraphics[width=3em]{figs/posts_joint.png}};
            \node[draw=none, below=0 of out1]{\footnotesize $v_t^{\spkrA}$};
            \node[draw=none, right=0.1em of out1, fill=white,fill opacity=1,text opacity=1, rounded corners=0.05cm, inner sep=0cm]{\footnotesize $v_t^{\spkrB}$};
            \node[coordinate, name=tarrow_st, below=0.5em of out4.south east] {};
            \node[coordinate, name=tarrow_en, right=0.5em of out6.south east] {};
            
            \node[draw=none, above=6mm of out1, text width=4.9cm, align=center]{joint output \\for~both speakers \\$p(v_t^{\spkrA},v_t^{\spkrB} | \y)$};
            
            \draw [->] (mix) -- (am);
            \draw [->] (am) -- (out4.west);
            \draw [->] (tarrow_st) -- node [right,xshift=-0.2em,yshift=-0.4em]{\footnotesize $t$} (tarrow_en);
            
            \node[below=1.4cm of am] {};
        \end{tikzpicture}
        \caption{Proposed model with joint output for both speakers.}
        \label{fig:joint_asr}
    \end{subfigure}
    \caption{Comparison of neural network architectures for single channel multi-talker speech recognition.}
\end{figure*}

\section{Introduction}
\label{sec:intro}
Although automatic speech recognition (ASR) systems today achieve remarkable performance, they degrade in presence of interference. Notably, recognizing speech in multi-talker environments is still a challenge, as demonstrated e.g. in recent CHiME evaluations~\cite{Barker2018,Watanabe2020}. The problem is even more pronounced in the single-channel scenario, where the recording of only one microphone is available. Significant amount of research addressing single-channel multi-talker speech recognition has been conducted in the past \cite{cooke2010monaural,QIAN20181,weng2015}, including early approaches using factorial generative models~\cite{Rennie2010,NIPS2006_67880768} and later neural network-based acoustic models~\cite{QIAN20181,weng2015}.

For multi-talker speech recognition, neural network-based systems are dominant nowadays \cite{QIAN20181,weng2015}. One of the most commonly used approaches is a multi-talker acoustic model trained with permutation invariant training (PIT)~\cite{kolbek2017,QIAN20181}.  In this case, the neural network constituting the acoustic model has separate outputs predicting the senone state posteriors for each speaker. The ambiguity in the order of speakers at the outputs and the labels is addressed by considering all possible permutations of the outputs during the training. In the inference time, the ASR decoder is applied separately on each sequence of posteriors of each speaker. PIT-ASR systems have shown to be promising for multi-talker ASR even for complex large vocabulary tasks \cite{QIAN20181,chang2018monaural}.

However, in the PIT-ASR approach, we expect the acoustic model to fully solve the ``separation'' of the speakers, i.e. fully attribute parts of the mixed speech signal to the different outputs of the network. This separation is based only on the acoustic signal and without higher level language information. This may be challenging especially when the voices of the speakers are very similar. While the decoder could help to solve the ``separation'' with the use of the dictionary and grammar, in the current scheme, there is no interaction between the decoders of the individual speakers. This can lead to duplicity where the same phoneme or word is attributed to multiple speakers~\cite{kanda2018hitachi}. In this paper we investigate a principled way to overcome this issue by \textit{decoding all speakers jointly}.

In pre-deep learning works, joint decoding of multiple speakers was proposed in the framework of factorial GMM-HMM (f-GMM-HMM)~\cite{Rennie2010}. This model showed success at recognizing speech in the monaural speech separation and recognition challenge~\cite{cooke2010monaural}, where it even outperformed human listeners~\cite{hershey2010super}. In the model, features of each speaker are modeled by GMM-HMMs with states corresponding to parts of speech (such as phonemes). Combining the per-speaker models using a speaker-interaction model gives rise to the f-GMM-HMM which can explain the mixed speech of multiple speakers. By doing inference of the most probable sequence of states, we can transcribe the speech of multiple speakers at once. This joint decoding of all speakers can take full advantage of the available dictionary and grammar, which can significantly improve the resulting hypothesis. The f-GMM-HMM has been however applied only to very constrained tasks with known speakers and is difficult to scale up to open speaker case, in part because of the iterative acoustic inference caused by the interaction model~\cite{Rennie2010}. Besides, the GMM-HMM ASR systems have been long outperformed by neural network-based systems~\cite{hinton2012deep,QIAN20181}.

In this work, we propose to combine the advantage of neural network-based approaches with the joint decoding used in factorial models. To do this, we modify the neural network in the acoustic model to predict joint state posteriors of multiple speakers given the mixed speech. By doing this, we enable the neural network to provide uncertainty about the attribution of the parts of speech to the speakers, which can be later resolved in the decoding stage using higher level language information. The joint state posteriors can be leveraged by the joint decoder based on factorial HMM. To make the inference feasible, we employ the loopy belief propagation algorithm as proposed in f-GMM-HMM works \cite{Rennie2010,rennie2009hierarchical}. Contrary to f-GMM-HMM, our system can take the advantage of the strong modeling power of neural networks. The neural network also simplifies the acoustic inference. In a proof of concept experiment, we demonstrate that 1) decoding each speaker separately in the PIT-ASR approach is not optimal, and 2) using the factorial HMM framework in combination with the neural network acoustic model, we can improve the performance by joint decoding of multiple speakers. 

%
%

\section{Multi-talker speech recognition}
\label{sec:mtsr}

In this work, we assume that the observed single-channel signal $\y$ is a combination of speech $\x^{\spkrA}$ and $\x^{\spkrB}$ from two speakers\footnote{To simplify the derivation, we restrict ourselves to the two speaker case and discuss extension to more speakers in Section~\ref{sec:discussion}.} as
\begin{equation}
    y_t = x_t^\spkrA + x_t^{\spkrB}.
\end{equation}
The goal is to recognize the speech of both speakers in the mixture.
This can be expressed as
\begin{equation}
    \hat{\w}^\spkrA, \hat{\w}^\spkrB = \argmax_{\w^\spkrA, \w^\spkrB} p(\y | \w^\spkrA, \w^\spkrB) P(\w^\spkrA) P(\w^\spkrB),
\end{equation}
where $\w^\spkrA = [v_1^\spkrA, \dots, v_T^\spkrA]$ is a hidden state sequence (which is mapped to a word sequence) for speaker $\spkrA$, $p(\y | \w^\spkrA, \w^\spkrB)$ is a likelihood function of observed mixed speech and $P(\w^\spkrA)$ and $P(\w^\spkrB)$ are prior probabilities estimated by language model.

Conventional multi-talker approaches such as PIT-ASR decode each speaker~$k$ separately as $\hat{\w}^k = \argmax p(\y|\w^k) p(\w^k)$~\cite{kolbek2017, QIAN20181}.
This rule can be obtained by assuming conditional independence of the state sequences of the speakers given the observation, i.e. $p(\w^a,\w^b|\y) = p(\w^a|\y)p(\w^b|\y)$.
However, this is a strong assumption that may not be valid in practice. We propose instead to consider the dependencies that should eventually lead to better modeling.

In this paper, we focus on DNN-HMM hybrid systems, where a DNN predicts senone states posteriors and HMM is used to model state transitions. In the following, we briefly review the conventional scheme for PIT-ASR with separate decoding and then introduce our proposed joint decoding with factorial DNN-HMM.


\subsection{Conventional PIT-ASR with separate decoding}
\label{sec:conventional_asr}

In PIT-ASR, the acoustic model predicts the posterior probabilities for each speaker separately
\begin{equation}
    [p(v_t^{\spkrA}|y_t), p(v_t^{\spkrB}|y_t)] = f_{\dnn}(y_t),
    \label{eq:pit_asr}
\end{equation}
where $p(v_t^{\spkrA}|y_t)$ and $p(v_t^{\spkrA}|y_t)$ are posterior probabilities of hidden states given observed frame $y_t$, $f_{\dnn}(y_t)$ is the DNN forward function. Such model is depicted in Figure~\ref{fig:conventional_asr}. Then, the ASR decoding is performed for each speaker independently using e.g. Viterbi algorithm. Below, we briefly explain the decoding considering the case of speaker $\spkrA$.


The joint probability $p(\x^\spkrA, \w^\spkrA)$ of source $\x^\spkrA$ and sequence of hidden states $\w^\spkrA$ is modeled by HMM
\begin{align}
    p(\x^\spkrA, \w^\spkrA) &= \prod_t p(x_t^\spkrA | v_t^\spkrA) p(v_t^\spkrA | v_{t-1}^\spkrA)\\
    p(x^\spkrA_t | v^\spkrA_t) &= \frac{p(v^\spkrA_t|x^\spkrA_t) p(x_t^\spkrA)}{p(v^\spkrA_t)}\\
    p(v^\spkrA_t|x^\spkrA_t) &\approx p(v^\spkrA_t|y_t),
\end{align}
where $p(v^\spkrA_t|y_t)$ is a posterior modeled by neural network \eqref{eq:pit_asr}.
Note that the evidence $p(x_t^a)$ does not affect the decoding results, since it is constant. That is why we use pseudo-likelihoods $\p(x^\spkrA_t | v^\spkrA_t)$
\begin{equation}
    \p(x^\spkrA_t | v^\spkrA_t) = \frac{p(v^\spkrA_t|y_t)}{ p(v_t^\spkrA)}
\end{equation}
However, in this work, we found out that normalizing by
prior $p(v_t^\spkrA)$  does not really help either, so we kept $\p(x^\spkrA_t | v^\spkrA_t) \approx p(v^\spkrA_t|y_t)$.

Recognizing speech using an HMM involves finding the most likely sequence $\hat{\w}^\spkrA$ given observed data $\x^\spkrA$ (i.e. the maximum a posteriori (MAP) state sequence). The MAP state sequence is obtained by Viterbi algorithm, where messages $m(v_t)$ are defined as
\begin{align}
    m(v_{t+1}^\spkrA) = \max_{v_t^\spkrA}\, & p(v_{t+1}^\spkrA|v_t^\spkrA) m(v_t^\spkrA) \p(x_t^\spkrA|v_t^\spkrA)\\
    \Tilde{v}_t(v_{t+1}^\spkrA) = \argmax_{v_t^\spkrA}\,& p(v_{t+1}^\spkrA|v_t^\spkrA) m(v_t^\spkrA) \p(x_t^\spkrA|v_t^\spkrA).
\end{align}
The MAP state sequence $\hat{\w}  = [\hat{v}_1, \dots, \hat{v}_T]$ is recovered by backtracking $\hat{v}_t = \Tilde{v}_t(\hat{v}_{t+1})$, where $\hat{v}_T = \argmax_{v_T} m(v_T)$ initiates the recursion.

\subsection{Proposed PIT-ASR with joint decoding}
\label{sec:joint_asr}

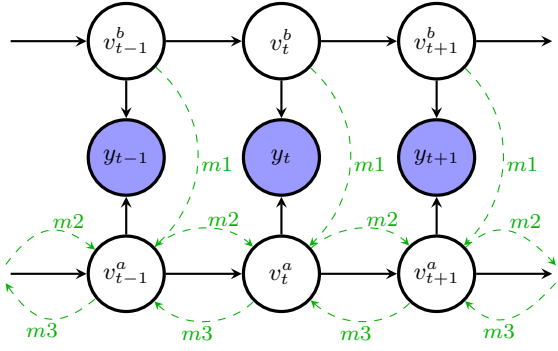
\begin{figure}[tb]
    \centering
    \begin{tikzpicture}[
            roundnode/.style={circle, draw=black!100, fill=white!100, very thick, minimum size=10mm},
            roundnode2/.style={circle, draw=white!100, fill=white!100, very thin, minimum size=0mm},
            >=stealth,
            roundnode3/.style={circle, draw=black!100, fill=myblue, very thick, minimum size=10mm},
            dashedarrow1/.style={dashed,->, mygreen},
            dashedarrow2/.style={dashed,<-, mygreen},
            simplearrow1/.style={thick, ->},
            simplearrow2/.style={thick, ->},
        ]
        \node[roundnode2]  (lefttop)                                            {};
        \node[roundnode]  (lefttopcircle)       [right=of lefttop]              {$v^\spkrB_{t-1}$};
        \node[roundnode]  (middletopcircle)     [right=of lefttopcircle]        {$v^\spkrB_{t}$};
        \node[roundnode]  (righttopcircle)      [right=of middletopcircle]      {$v^\spkrB_{t+1}$};
        \node[roundnode2]  (righttop)           [right=of righttopcircle]       {};
        
        \node[roundnode3]  (leftbottomy)         [below=0.5 of lefttopcircle]     {$y_{t-1}$};
        \node[roundnode3]  (middlebottomy)       [right=of leftbottomy]   {$y_{t}$};
        \node[roundnode3]  (rightbottomy)        [right=of middlebottomy]    {$y_{t+1}$};
        
        \node[roundnode]  (leftbottomcircle)    [below=0.5 of leftbottomy]           {$v^\spkrA_{t-1}$};
        \node[roundnode2]  (leftbottom)         [left=of leftbottomcircle]           {};
        \node[roundnode]  (middlebottomcircle)  [right=of leftbottomcircle]     {$v^\spkrA_{t}$};
        \node[roundnode]  (rightbottomcircle)   [right=of middlebottomcircle]   {$v^\spkrA_{t+1}$};
        \node[roundnode2]  (rightbottom)        [right=of rightbottomcircle]    {};
        
        \draw[simplearrow1] (lefttop.east) -- (lefttopcircle.west);
        \draw[simplearrow1] (lefttopcircle.east) -- (middletopcircle.west);
        \draw[simplearrow1] (middletopcircle.east) -- (righttopcircle.west);
        \draw[simplearrow1] (righttopcircle.east) -- (righttop.west);
        
        \draw[simplearrow1] (leftbottom.east) -- (leftbottomcircle.west);
        \draw[simplearrow1] (leftbottomcircle.east) -- (middlebottomcircle.west);
        \draw[simplearrow1] (middlebottomcircle.east) -- (rightbottomcircle.west);
        \draw[simplearrow1] (rightbottomcircle.east) -- (rightbottom.west);
        
        \draw[simplearrow2] (lefttopcircle) -- (leftbottomy);
        \draw[simplearrow2] (leftbottomcircle) -- (leftbottomy);
        
        \draw[simplearrow2] (middletopcircle) -- (middlebottomy);
        \draw[simplearrow2] (middlebottomcircle) -- (middlebottomy);
        
        \draw[simplearrow2] (righttopcircle) -- (rightbottomy);
        \draw[simplearrow2] (rightbottomcircle) -- (rightbottomy);

        \draw[dashedarrow1] (lefttopcircle) .. controls(2.8,-1) and (2.8,-2) .. node [right,xshift=-0.2em,yshift=-0.4em]{\footnotesize{$m1$}}(leftbottomcircle);
        \draw[dashedarrow1] (middletopcircle) .. controls(4.8,-1) and (4.8,-2) .. node [right,xshift=-0.2em,yshift=-0.4em]{\footnotesize{$m1$}}(middlebottomcircle);
        \draw[dashedarrow1] (righttopcircle) .. controls(6.8,-1) and (6.8,-2) .. node [right,xshift=-0.2em,yshift=-0.4em]{\footnotesize{$m1$}}(rightbottomcircle);
        
        \draw[dashedarrow1] (leftbottom) .. controls(0.5,-2.4) and (0.8, -2.4) .. node [right,xshift=-0.2em,yshift=+0.4em]{\footnotesize{$m2$}}(leftbottomcircle);
        \draw[dashedarrow1] (leftbottomcircle) .. controls(2.4,-2.4) and (3, -2.4) .. node [right,xshift=-0.2em,yshift=+0.4em]{\footnotesize{$m2$}}(middlebottomcircle);
        \draw[dashedarrow1] (middlebottomcircle) .. controls(4.5,-2.4) and (5.1, -2.4) .. node [right,xshift=-0.2em,yshift=+0.4em]{\footnotesize{$m2$}}(rightbottomcircle);
        \draw[dashedarrow1] (rightbottomcircle) .. controls(6.5,-2.4) and (6.8,-2.4) .. node [right,xshift=-0.2em,yshift=+0.4em]{\footnotesize{$m2$}}(rightbottom);
        
        \draw[dashedarrow2] (leftbottom) .. controls(0.5,-3.8) and (0.8, -3.8) .. node [right,xshift=-1.3em,yshift=-0.5em]{\footnotesize{$m3$}}(leftbottomcircle);
        \draw[dashedarrow2] (leftbottomcircle) .. controls(2.4,-3.8) and (3, -3.8) .. node [right,xshift=-1.3em,yshift=-0.5em]{\footnotesize{$m3$}}(middlebottomcircle);
        \draw[dashedarrow2] (middlebottomcircle) .. controls(4.5,-3.8) and (5.1, -3.8) .. node [right,xshift=-1.3em,yshift=-0.5em]{\footnotesize{$m3$}}(rightbottomcircle);
        \draw[dashedarrow2] (rightbottomcircle) .. controls(6.5,-3.8) and (6.8,-3.8) .. node [right,xshift=-1.3em,yshift=-0.5em]{\footnotesize{$m3$}}(rightbottom);
        
    \end{tikzpicture}
    \caption{Factorial HMM for speaker~$\spkrA$ and $\spkrB$ with observed sequence $\y$ and hidden state sequences $\w^\spkrA$ and $\w^\spkrB$. Green dashed lines show message passing in loopy belief propagation.}
    \label{fig:fhmm}
\end{figure}

To take account of the dependencies, we model the joint probability of mixed speech $\y$ and hidden state sequences $\w^{\spkrA}, \w^{\spkrB}$ in Factorial HMM framework~\cite{Ghahramani1997}.
Figure~\ref{fig:fhmm} shows dynamic Bayesian network representing this model, where
the joint probability is
\begin{equation}
    p(\y, \w^{\spkrA}, \w^{\spkrB}) = \prod_t p(y_t | v_t^{\spkrA}, v_t^{\spkrB}) p(v_t^{\spkrA}|v_{t-1}^{\spkrA}) p(v_t^{\spkrB}|v_{t-1}^{\spkrB}),
\end{equation}
where $ p(y_t | v_t^{\spkrA}, v_t^{\spkrB}) $ is derived from a neural network that predicts the posterior probabilities of a tuple of states $(v_t^{\spkrA}, v_t^{\spkrB})$ as  
\begin{equation}
    p(v_t^{\spkrA}, v_t^{\spkrB} | y_t) = g_{\dnn}(y_t),
\end{equation}
where $g_{\dnn}(y_t)$ is a forward function of proposed neural network shown in Figure~\ref{fig:joint_asr} (also referred as joint network in follow-up text).
Similar to the separate decoding, we use pseudo-likelihood $\p(y_t | v_t^{\spkrA}, v_t^{\spkrB}) \approx p(v_t^{\spkrA}, v_t^{\spkrB} | y_t)$.

Note that our model is similar to Factorial HMM proposed in~\cite{Rennie2010}, where the authors model sources $\x^\spkrA$ and $\x^\spkrB$ explicitly. They introduce  interaction function $p(y_t | \x^\spkrA, \x^\spkrB)$, which models the coupling. However, we instead use a more direct and simpler approach where a DNN is trained to directly predict hidden state tuples given mixed speech $\y$.

Standard Viterbi algorithm can be used to exactly infer the MAP hidden state sequences in FHMM. However, its time complexity is $O(TKV^{K+1})$ for $K$ speakers and HMM with $V$ states~\cite{Rennie2010} (i.e. it scales exponentially w.r.t. number of speakers).
Fortunately, there are algorithms, that can approximate the inference in a more efficient way (e.g. variational methods~\cite{Ghahramani1997} or loopy belief propagation~\cite{Rennie2010}). In this work we use the loopy belief propagation (LBP).

In LBP framework, messages are passed between variables, which share common factors~\cite{Bishop2006} according to a predefined schedule. Our message passing schedule was inspired by~\cite{Rennie2010} and has the following form for one speaker:
\begin{align*}
    \Tilde{p}(y_t | v_t^{\spkrA}) &= \max_{v_{t}^{\spkrB}}  \p(y_t | v_t^{\spkrA}, v_t^{\spkrB}) \Tilde{p}_{fw}(v_t^{\spkrB})     \Tilde{p}_{bw}(v_t^{\spkrB})          \tag{m1}\\
    \Tilde{p}_{fw}(v_t^{\spkrA})  &= \max_{v_{t-1}^{\spkrA}} p(v_t^{\spkrA} | v_{t-1}^{\spkrA}) \Tilde{p}_{fw}(v_{t-1}^{\spkrA}) \Tilde{p}(y_{t-1} | v_{t-1}^{\spkrA}) \tag{m2}\\
    \Tilde{p}_{bw}(v_t^{\spkrA})  &= \max_{v_{t+1}^{\spkrA}} p(v_{t+1}^{\spkrA} | v_t^{\spkrA}) \Tilde{p}_{bw}(v_{t+1}^{\spkrA}) \Tilde{p}(y_{t+1} | v_{t+1}^{\spkrA}) \tag{m3}\\
\end{align*}
The message passing is also depicted in Figure~\ref{fig:fhmm}. Green dashed lines depict the message passing for speaker~$\spkrA$, while messages of speaker~$\spkrB$~are fixed.

When all messages for the first speaker are computed, the process is repeated for the next speaker, while messages of other speakers are fixed.
Similar to Viterbi, we store also maximizing arguments $\Tilde{v}_t$ for all messages. These are used to recover the MAP state sequences in the manner analogous to the Viterbi algorithm. 
The whole process is repeated until it converges or some number of iterations is reached. Although there is no guarantee that the LBP finds the global MAP configuration, it is known to work well in practice \cite{weiss2001}.  The time complexity of proposed LBP inference is $O(TKV^2)$, and thus it scales linearly w.r.t. number of speakers.

%
%

\section{Experiments}
\label{sec:experiments}


\subsection{TIDIGITS 2 mixture data}
\label{sec:dataset}

To evaluate the performance of proposed architectures, we conducted experiments on artificially mixed TIDIGITS dataset~\cite{leonard1984}. Each mixture consists of exactly two speakers randomly selected from original TIDIGITS. Each utterance is a sequence of pronounced digits, the length may vary from 1 to 7 digits. The training data contains $6\,328$ different pairs of speakers, $52.5$ hours of speech. The evaluation data comprises of $4\,821$ different speaker pairs, in total $5.3$ hours of audio.
More details about the dataset are available on GitHub\footnote{\url{https://github.com/MartinKocour/TIDIGITS_mix}}.
We use this simple dataset to allow for easy analysis of the method and fast experimentation. For discussion on scaling up to larger data, see Section~\ref{sec:discussion}.



\subsection{Acoustic model}

Both the PIT-ASR with separate decoding and the proposed ASR with joint decoding employ a DNN-based acoustic model. Both models are trained in PyTorch~\cite{PyTorch} using permutation invariant training (PIT)\cite{kolbek2017,QIAN20181} with cross entropy (CE) loss function. We can use PIT-CE objective function for both approaches, since, for every frame, both neural networks predict probabilities of being in HMM state $v_t^\spkrA$ and $v_t^\spkrB$ although, the state probabilities are modeled in different manner.

Both networks shares the same architecture: TDNN~\cite{peddinti15b_tdnn} layers with 384 units followed by batch normalization and ReLu activation function. We compared neural networks with 5 or 10 TDNN layers. The neural network for separate decoding contains 2 output layers with size of 62 each. On the other hand, the neural network used with joint decoding consist of single output layer with the size of $62\times62$, i.e. $3844$. Note that the size of joint network output layer grows exponentially w.r.t. number of speakers. We discuss how this issue can be addressed in Section~\ref{sec:discussion}. The input consists of 40-dimensional MFCCs.

The frame-level labels (i.e. state index for each time frame) for each source $l_t^\spkrA$ and $l_t^\spkrB$ are obtained by applying the force-alignment on original sources $\x^\spkrA$ and $\x^\spkrB$.  We generate the labels by mono-phone GMM-HMM models, taken from Kaldi~\cite{Kaldi} recipe for TIDIGITS dataset.
For the conventional separate PIT model, the CE loss target probabilities
are $[\hat{p}(v_t^{\spkrA}|y_t), \hat{p} (v_t^{\spkrB}|y_t)]$ where $\hat{p}(v_t^{\spkrA}=i|y_t) = \delta_{l_t^{\spkrA},i}$, 
where $\delta_{l_t^{\spkrA},i} = 1$ if training label $l_t^{\spkrA} = i$ and is zero otherwise.
For the joint decoding model, we derive the target probabilities as $ \hat{p}(v_t^{\spkrA}=i, v_t^{\spkrB}=j | y_t) = \delta_{l_t^{\spkrA},i} \delta_{l_t^{\spkrB},j} $.

\subsection{Decoding network}

The recognition network is similar to WFST used in Kaldi TIDIGITS setup~\footnote{The recipe is available on GitHub \url{https://github.com/kaldi-asr/kaldi/tree/master/egs/tidigits}} for monophone GMM-HMM ASR system. The network is based on unigram LM, where each word, i.e. digit, is equally likely. Pronunciations are taken from CMU Dictionary\footnote{The dictionary is avaialable on \url{https://github.com/cmusphinx/cmudict}}. We model silence with 5 HMM states and the other 19 phones with 3 HMM states. The HMM states do not share emission probabilities, i.e. we use 62 distinct PDFs in total. The network is implemented in Julia~\cite{Julia} using the MarkovModels toolkit~\footnote{The toolkit is available on GitHub \url{https://github.com/lucasondel/MarkovModels.jl}}.

%
%

\section{Results}
\label{sec:results}

We evaluate the speech recognition systems in all experiments using Word error rate (WER). To pair the multi-talker hypotheses with the references, we use the oracle permutation with minimal WER.

\subsection{Comparison of joint and separate decoding}

We evaluate the performance of separate and joint decoding on $5.3$ hours of 2-talker mixed speech.
We use two different sizes of TDNN with 5 layers and 10 layers. In Table~\ref{tab:main_results}, we compare the results obtained with both models using \textit{separate} and \textit{joint} decoding. 
For a more fair comparison, we include third \textit{separate-marginal} method marked with $\Sigma$-symbol, which combines the AM predicting joint posteriors (Sec.~\ref{sec:joint_asr}) with separate decoding (Sec.~\ref{sec:conventional_asr}). For this, we marginalize the joint posteriors for each speaker as $p(v^{\spkrA}_t|y_t) = \sum_{v^{\spkrB}_t} p(v^{\spkrA}_t, v^{\spkrB}_t | y_t)$. This allows us to separately evaluate the benefit of the joint posterior output (which also induces increased number of parameters) and the benefit of the joint decoding itself.

By comparing the results of \textit{separate} decoding in rows 1 and 2 for the smaller model, and 3 and 4 for the larger model, we can see the advantage of the joint posteriors on the output of the acoustic model. More importantly, comparing the \textit{separate-marginal} decoding with the \textit{joint} decoding, we can see improvements for both smaller and larger models, demonstrating the benefit of decoding both speakers jointly in the factorial framework.

We also compare the performance of Kaldi decoder with our implementation of separate decoding.
The results slightly differ when marginalized joint posteriors are used, which we believe is caused by pruning. In our approach, we do not use any kind of pruning.




\begin{table}[tb]
\centering
\caption{Comparison of separate and joint decoding in terms of WER. The results marked with $\Sigma$-symbol are computed from marginalized joint posteriors.}
\begin{tabular}{@{}lrr|ll|l@{}}
\toprule
Arch     & Output       & \#params   & Separate         & Joint & Kaldi\\
         & {[}dim{]}    &            &                  &       &      \\\midrule
5L-TDNN  & {[}62, 62{]} & $1.9$\,M   & $26.09$          & -     & $25.99$\\
5L-TDNN  & {[}3844{]}   & $3.3$\,M   & $17.55^\Sigma$   & $15.79$ & $17.83^\Sigma$\\
10L-TDNN & {[}62, 62{]} & $4.1$\,M   & $18.68$.         & -     & $18.66$\\
10L-TDNN & {[}3844{]}   & $5.5$\,M   & $16.36^\Sigma$   & $\textbf{14.70}$ & $16.97^\Sigma$\\
\bottomrule
\end{tabular}
\label{tab:main_results}
\end{table}

\subsection{Analysis of 2-talker speech recognition results}

We further analyzed the WER differences between separate and joint decoder on the speech mixtures of same or opposite genders.
The results are shown in Table~\ref{tab:gender_results}. We observe that the proposed joint decoding greatly improves performance on same gender mixtures, where~it may be particularly difficult to distinguish speakers. We hypothesize that thanks to employing dictionary and grammar (in recognition network) to ``separate'' speech, the proposed method can better discriminate the speakers than the separate decoding scheme performing ``separation'' only based on the acoustic information.

\begin{table}[tb]
\centering
\caption{Comparison of joint and separate decoder on the mixed speech of two female speakers (F + F), two male speakers (M + M), same (F + F $\lor$ M + M) and opposite gender speakers (F + M).}
\begin{tabular}{r|cc}
\toprule
Genders & Separate {[}\%WER{]} & Joint {[}\%WER{]} \\\midrule
F\,+\,F     &     $30.54$        &      $21.45$      \\
M\,+\,M     &     $32.61$        &      $27.12$       \\\midrule
same        &     $31.56$        &      $24.26$       \\
opposite    &     $6.17$         &      $5.42$       \\
\bottomrule         
\end{tabular}
\label{tab:gender_results}
\end{table}

\section{Discussion and Future work}
\label{sec:discussion}


In this study we analyzed the performance on rather simple task as a proof-of-concept of the joint decoding scheme. The results suggest that considering joint decoding could greatly improve recognition performance of multi-speaker ASR. These results come at the expense of increased computational complexity. For example, in our experiments we observed that the real time factor of the decoding on a CPU increased from approximately $0.3$ with separate decoding to $0.5$ with joint decoding. However, the decoding time would further increase when dealing with larger vocabulary tasks where the number of senones and the computational graph is larger. In future work, we plan to reduce the computational complexity by investigating factorized model for the joint posteriors and implementing fast decoder using pruning, sparse matrices and GPU-based computations\cite{ondel2021}, which would allow tackling larger tasks.

\section{Conclusion}

This work investigates the potential of joint-decoding for multi-talker speech recognition, in contrast with separate decoding in conventional approaches.
The proposed method is based on early multi-talker speech recognition works, which use generative factorial approaches. Our work extends this ideas by replacing the acoustic model with DNN. The proof-of-concept experiments showed that the approach has potential to improve performance in challenging conditions where it may be difficult to achieve high separation by simply relying on the acoustic information (e.g. potentially when dealing with noisy and reverberant mixtures). We hope that the findings of this study will promote the importance of joint decoding for future research in multi-talker ASR.


\vfill\pagebreak


\bibliographystyle{IEEEtran}

\bibliography{refs}

\end{document}